\documentclass[aps,prl,twocolumn,superscriptaddress,preprintnumbers,longbibliography]{revtex4-2}
\usepackage{amsmath}
\usepackage{amssymb}
\usepackage{graphicx}
\usepackage{gensymb}
\usepackage{xcolor}
\usepackage{float}
\usepackage{physics}
\usepackage{siunitx}
\usepackage{comment}
\usepackage[colorlinks = true, citecolor = blue, linkcolor = blue, urlcolor = blue]{hyperref}
\usepackage{placeins}
\usepackage[normalem]{ulem}
\newcommand{\bea}{\begin{eqnarray}}
\newcommand{\eea}{\end{eqnarray}}
\newcommand{\nn}{\nonumber}

\begin{document}
\newcommand{\DP}[1]{\textbf{\color{green}DP: #1}}

\title{One-point charge correlator as a probe for the odderon}

\author{Haotian Cao}
\email{haotiao.cao@northwestern.edu}
 \affiliation{Department of Physics and Astronomy, Northwestern University, Evanston, Illinois 60208, USA}
\affiliation{Center for Frontiers in Nuclear Science, Stony Brook University, Stony Brook, NY 11794, USA}

\author{Zhong-Bo Kang}
\email{zkang@physics.ucla.edu}
\affiliation{Department of Physics and Astronomy, University of California, Los Angeles, CA 90095, USA}
\affiliation{Mani L. Bhaumik Institute for Theoretical Physics, University of California, Los Angeles, CA 90095, USA}
\affiliation{Center for Frontiers in Nuclear Science, Stony Brook University, Stony Brook, NY 11794, USA}

\author{Diego Padilla}
\email{dpadi022@g.ucla.edu}
\affiliation{Department of Physics and Astronomy, University of California, Los Angeles, CA 90095, USA}
\affiliation{Mani L. Bhaumik Institute for Theoretical Physics, University of California, Los Angeles, CA 90095, USA}

\author{Jani Penttala}
\email{jani.penttala@ijclab.in2p3.fr}
\affiliation{Université Paris-Saclay, CNRS/IN2P3, IJCLab, 91405 Orsay, France}
\affiliation{Department of Physics and Astronomy, University of California, Los Angeles, CA 90095, USA}
\affiliation{Mani L. Bhaumik Institute for Theoretical Physics, University of California, Los Angeles, CA 90095, USA}

\begin{abstract}
We propose the one-point charge correlator (OPCC) in transversely polarized deep inelastic scattering as a new probe of the spin-dependent odderon. The OPCC is an infrared and collinear safe observable constructed solely from the charge and angular information of final-state charged particles. Because the charge weight is odd under charge conjugation, the contribution of the C-even pomeron to the OPCC vanishes identically in the small-$x$ eikonal limit, whereas the contribution of the C-odd spin-dependent odderon survives. We define a transverse single-spin asymmetry by normalizing the spin-dependent OPCC to the unpolarized charged-hadron one-point energy correlator. At small $x$, this asymmetry reduces to a ratio of spin-dependent odderon and pomeron contributions. We provide illustrative estimates of this asymmetry for Electron-Ion Collider kinematics within the small-$x$ dipole framework with Balitsky--Kovchegov evolution to guide experimental studies. Dedicated measurements of the OPCC asymmetry with a transversely polarized proton beam would provide new quantitative constraints on the currently unconstrained spin-dependent odderon.
\end{abstract}

\maketitle 


{\it \textbf{Introduction.}} 
The odderon, a C-odd color-singlet $t$-channel exchange~\cite{Lukaszuk:1973nt,Joynson:1975az,Ewerz:2003xi}, has remained notoriously elusive in QCD phenomenology~\cite{H1:2002ckt,Olsson:2001nm}. Experimental evidence for it has only recently been reported from the comparison of $pp$ and $p\bar{p}$ elastic scattering~\cite{D0:2012erd,TOTEM:2018psk,D0:2020tig}. The interpretation of this evidence relies on modeling the $t$ dependence of the elastic amplitudes, and quantitative constraints on the odderon from independent processes are still lacking.

In the small-$x$ limit of deep inelastic scattering (DIS) ~\cite{Gelis:2010nm,Iancu:2003xm,McLerran:1994vd}, 
the interaction between the projectile and the target is described by the dipole picture~\cite{Nikolaev:1990ja,Nikolaev:1991et,Mueller:1993rr,Mueller:1994gb,Mueller:1994jq}, in which the dipole amplitude resums the eikonal scattering of a quark--antiquark dipole off the gluon field of the target. The real part of the dipole amplitude is even under charge conjugation and is identified with the pomeron, while the imaginary part is odd and is identified with the odderon~\cite{Hatta:2005as}. 
This C-odd exchange can be isolated by considering particular final states~\cite{Lorce:2013pza}. It has been suggested that transverse spin-asymmetries~\cite{Zhou:2013gsa,Hatta:2026iry}, charge asymmetries~\cite{Brodsky:1999mz,Hagler:2002cq} and exclusive production of C-even mesons in DIS~\cite{Czyzewski:1996bv,Engel:1997cga,Dumitru:2019qec,Benic:2023ybl,Benic:2024pqe,Benic:2024fbf} are especially sensitive to the odderon. 
A particularly interesting projection of the dipole amplitude is the spin-dependent odderon, which is related to the dipole gluon Sivers function in the small-$x$ limit~\cite{Zhou:2013gsa,Boer:2015pni,Szymanowski:2016mbq,Yao:2018vcg,Boussarie:2019vmk,Benic:2026nnf}. Along the same lines, it has been shown that the small-$x$ sea-quark Sivers function can also be expressed in terms of the spin-dependent odderon~\cite{Dong:2018wsp,Kovchegov:2021iyc,Bhattacharya:2025bqa}, enabling new phenomenological probes.

Recent developments in energy correlators and related observables~\cite{Basham:1978zq,Basham:1978bw,Basham:1977iq,Ore:1979ry,Sveshnikov:1995vi,Korchemsky:1997sy,Korchemsky:1999kt,Belitsky:2001ij,Lee:2006nr,Hofman:2008ar,Moult:2025nhu} have opened a promising avenue for probing small-$x$ physics~\cite{Liu:2022wop,Kang:2023oqj,Mantysaari:2025mht,Kang:2025vjk,Kang:2026hig, Mantysaari:2026zte}. 
In this context, the spin-dependent nucleon energy correlator~\cite{Mantysaari:2025mht} and the transverse energy--energy correlator~\cite{Bhattacharya:2025bqa} have been proposed as novel observables sensitive to the spin-dependent odderon. In both cases, however, the Sivers signal cancels between the quark and antiquark channels at small $x$ unless the final state is restricted to hadrons of a given charge sign.   The one-point charge correlators~\cite{Hofman:2008ar,Belitsky:2013xxa,Belitsky:2013bja,Chicherin:2020azt,Riembau:2024tom,Li:2026fhd,Zhang:2026vdo,Zhang:2026emt,Zhang:2026esd}, which replace the energy weight with the electric charge, flip the charge conjugation parity of the final-state weight and therefore provide a natural way to turn this cancellation into an enhancement. 

In this Letter, we propose a novel approach to probe the odderon in $ep$ collisions using the recently proposed one-point charge correlator (OPCC)~\cite{Cao:2026fzq,Cao:2026kcg}. This observable is defined as a charge-weighted cross section that is differential in the angle of the detected hadrons in the final state. We show that, at small $x$ and to eikonal accuracy, the C-even pomeron contribution to the OPCC vanishes identically, while the C-odd spin-dependent odderon contribution survives and is doubled by the addition of the antiquark channel. We define a transverse single-spin asymmetry by normalizing the spin-dependent OPCC to the unpolarized charged-hadron one-point energy correlator. At small $x$, this asymmetry becomes a ratio of spin-dependent odderon and pomeron convolutions and can be predicted within the dipole framework with fragmentation input. Experimentally, both correlators use charged-particle tracking, without requiring jet reconstruction or hadron identification.

{\it \textbf{Observables and factorization.}} 
In this Letter, we consider DIS of an unpolarized electron off a transversely polarized proton, $e\,p^\uparrow \to e\, h + X$, in the Breit frame, and define the family of weighted one-point correlators
\begin{align} \label{eq:OPCC-w}
\Sigma_{w}(\vec n,x,Q^2) = \sum_h \int d\sigma_{ep^\uparrow \to e h + X} \, w_h \, \delta(\vec n - \vec n_h) \,,
\end{align}
where $\vec n_h$ is the direction of the outgoing hadron, $\vec n$ is the detector direction specified by the polar and azimuthal angles $(\theta,\varphi)$ in the Breit frame, $x$ is the Bjorken variable, and $Q^2$ is the photon virtuality. The polar angle $\theta$ is measured relative to the incoming proton direction. Here $w_h \equiv w(z_h,h)$ is the measurement weight. The OPCC~\cite{Cao:2026fzq,Cao:2026kcg} corresponds to the charge weight $w_h = Q_h$, the electric charge of the hadron in units of $e$, and we denote it by $w={\cal Q}$. The energy weight $w_h = z_h \equiv P\cdot p_h/P\cdot q$, with $P$, $p_h$, and $q$ the four-momenta of the proton, hadron, and virtual photon, defines the one-point energy correlator~\cite{Basham:1978bw,Basham:1978zq}, which we denote by $w = E$. In both cases the sum runs over {\it charged hadrons} only, so that the observables are measured with tracking alone~\cite{Zhang:2026vdo,Electron-PositronAlliance:2025fhk}.  

\begin{figure}[t]
\begin{center}
\includegraphics[width=\columnwidth]{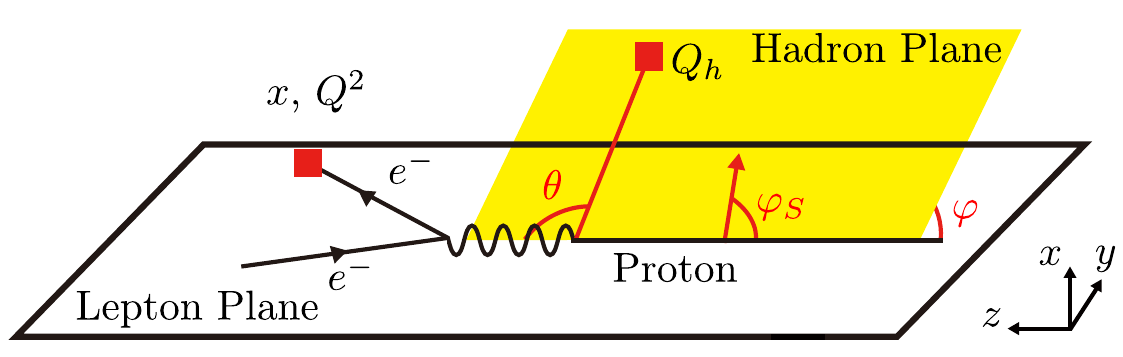}
\caption{Schematic view of the OPCC measurement in transversely polarized DIS, with the relevant angles defined in the Breit frame.}
\label{fg:obs}
\end{center}
\end{figure}

The observable is illustrated schematically in Fig.~\ref{fg:obs}. We consider hadrons emitted nearly opposite to the incoming proton, corresponding to the back-to-back region ($\theta\to\pi$), and define $\bar\theta \equiv \pi - \theta \ll 1$. The corresponding transverse recoil variable is $q_T \equiv |\vec{P}_{hT}|/z_h\simeq Q\bar\theta/2$, where $\vec{P}_{hT}$ is the hadron transverse momentum. For a transversely polarized proton with spin vector $\vec S_T$ at azimuthal angle $\varphi_S$, the correlators decompose as
\begin{align}
\Sigma_{w} (\vec{n},x,Q^2)
=
 F_{UU}^{w} +  |\vec S_T| \sin(\varphi-\varphi_S)\,
F_{UT}^{w} + \dots \,,
  \label{eq:aut}
\end{align}
where the ellipsis denotes other angular structures not relevant here, and we follow the Trento conventions~\cite{Bacchetta:2004jz} for the sign of the Sivers term. Throughout, $\Sigma_w$ is differential in $d\theta\, d\varphi$, as in Ref.~\cite{Cao:2026kcg}.

At leading power in ${\bar \theta}^2$, the unpolarized and polarized structure functions factorize in the transverse-momentum-dependent (TMD) framework~\cite{Collins:2011zzd,Boussarie:2023izj,Cao:2026kcg,Cao:2026fzq}   
\begin{align}
\label{eq:TMDunpolfac}
    F_{UU}^{w}& = \sigma_0
        H(Q^2,\mu)   \frac{{\bar \theta} Q^2}{4}  \int_0^{\infty} \frac{b\, db}{2\pi}J_0\left( b \frac{Q{\bar \theta}}{2}  \right)\nn \\
         \times & \sum_{i=q,\bar q} \,e_i^2\,
         J_{i,w}(b,\mu,\zeta) \,
         f_{i/p} (x,b,\mu,\zeta) \, ,
\\
\label{eq:TMDpolfac}
    F_{UT}^{w}& = \sigma_0
        H(Q^2,\mu)  \frac{{\bar \theta}Q^2}{4}  \int_0^{\infty} \frac{b^2\, db}{4\pi}J_1\left( b \frac{Q{\bar \theta}}{2}  \right)\nn \\
         \times & \sum_{i=q,\bar q}\,e_i^2\,
         J_{i,w}(b,\mu,\zeta) \,
         M_p f_{1T,i/p}^{\perp}(x,b,\mu,\zeta) \, ,
\end{align} 
where $M_p$ is the proton mass, $J_0$ and $J_1$ are Bessel functions of the first kind, $\sigma_0$ is the partonic Born cross section, $H$ is the hard function, $e_i$ is the quark electric charge, and the sum runs over quark and antiquark flavors. The functions $f_{i/p}$ and $f_{1T,i/p}^{\perp}$ are the unpolarized quark TMD parton distribution function (PDF) and the quark Sivers function, evaluated at the renormalization scale $\mu$ and the Collins--Soper scale $\zeta$. The jet function $J_{i,w}$ is related to the TMD fragmentation functions (FFs) $D_{h/i}$ as follows:
\begin{align}
\label{eq:Jetfunc}
J_{i,{w}}(b, \mu, \zeta)  = \sum_h \int_{z_{\rm min}}^1 dz_h  \,\omega(z_h, h)\, D_{h/i}(z_h, b, \mu, \zeta) \,.
\end{align}
Here $z_{\rm min}=0$ corresponds to the unrestricted integral. A finite $z_{\rm min}$ can reflect the lower limit adopted for the FF parametrization or an experimental acceptance cut. The dependence of the jet function on the cutoff is left implicit. For the charge weight $Q_h$, this is the same as the charge--charge correlation jet function of Ref.~\cite{Monni:2025zyv}. For the energy weight $z_h$, it corresponds to the energy-energy correlation jet function of Ref.~\cite{Moult:2018jzp}, with the hadronic sum restricted to charged hadrons. Under charge conjugation, $Q_h \to -Q_h$ while $z_h$ is unchanged, so that
\begin{align}
\label{eq:Cjet}
J_{\bar q,{\cal Q}} = - J_{q,{\cal Q}}\,, \qquad J_{\bar q,E} = + J_{q,E}\,.
\end{align}

We define the OPCC Sivers asymmetry as
\begin{align}
\label{eq:siversdef}
A_{\rm Sivers}(\theta,x) \equiv \frac{F^{\cal Q}_{UT}}{F^{E}_{UU}}\,,
\end{align}
i.e., the spin-dependent charge flow normalized to the charge-blind energy flow in the same angular bin. Both structure functions share $\sigma_0$, $H$, the $\bar\theta$ prefactor, and the lepton tensor, so these cancel in the ratio. Because the numerator and denominator use different weights, $A_{\rm Sivers}$ is not bounded by unity.

{\it \textbf{Charge conjugation and the small-$x$ limit.}} 
In the high-energy limit ($x\ll1$) the quark content of the proton is dominated by sea quarks, which are sourced by the enhanced gluon content of the proton. To eikonal accuracy, the sea-quark TMDs then satisfy~\cite{Bhattacharya:2025bqa,Dong:2018wsp,Tong:2022zwp,Marquet:2009ca,Kovchegov:2021iyc}
\begin{align}
\label{eq:Ctmd}
f_{\bar q/p} = + f_{q/p}\,, \qquad f^\perp_{1T,\bar q/p} = - f^\perp_{1T,q/p}\,,
\end{align}
reflecting the fact that the unpolarized sea is generated by the C-even pomeron while the sea-quark Sivers function is generated by the C-odd odderon, as we make explicit below.

Combining Eqs.~\eqref{eq:Cjet} and~\eqref{eq:Ctmd}, the flavor sums in Eqs.~\eqref{eq:TMDunpolfac} and~\eqref{eq:TMDpolfac} project onto definite C-parity combinations. Writing the sum over $i=q,\bar q$ as a sum over quark flavors $q=u,d,s$ and abbreviating $f_q \equiv f_{q/p}$,
\begin{align}
\label{eq:projections}
F_{UU}^{\cal Q} &\propto \sum_q e_q^2\, J_{q,{\cal Q}}\,\big(f_{q} - f_{\bar q}\big) \;\to\; 0\,, \nn\\
F_{UT}^{\cal Q} &\propto \sum_q e_q^2\, J_{q,{\cal Q}}\,\big(f^\perp_{1T,q} - f^\perp_{1T,\bar q}\big) \;\to\; 2\sum_q e_q^2\, J_{q,{\cal Q}}\, f^\perp_{1T,q}\,, \nn\\
F_{UT}^{E} &\propto \sum_q e_q^2\, J_{q,E}\,\big(f^\perp_{1T,q} + f^\perp_{1T,\bar q}\big) \;\to\; 0\,, \nn\\
F_{UU}^{E} &\propto \sum_q e_q^2\, J_{q,E}\,\big(f_{q} + f_{\bar q}\big) \;\to\; 2\sum_q e_q^2\, J_{q,E}\, f_{q}\,,
\end{align}
where the arrows denote the eikonal small-$x$ limit. A central prediction is that the C-even pomeron contribution to the unpolarized OPCC cancels identically between the quark and antiquark channels, yielding $F_{UU}^{\cal Q}=0$ in this limit. In contrast, the charge weight compensates the sign reversal of the odderon-generated sea-quark Sivers function, so that the spin-dependent OPCC $F_{UT}^{\cal Q}$ survives and is doubled by the addition of the antiquark channel. This distinguishes the OPCC from energy-weighted probes, for which the spin-dependent contributions cancel, giving $F_{UT}^{E}=0$~\cite{Bhattacharya:2025bqa,Mantysaari:2025mht}. The unpolarized energy correlator $F_{UU}^{E}$ remains nonzero and provides the normalization in Eq.~\eqref{eq:siversdef}.

At small $x$, the sea-quark TMDs at the initial scale $(\mu_i, \zeta_i)$ can be expressed in terms of the dipole amplitude~\cite{Bhattacharya:2025bqa,Dong:2018wsp,Tong:2022zwp,Marquet:2009ca,Kovchegov:2021iyc}. Their evolution to the final scales $(\mu, \zeta)$ is governed by TMD evolution, which resums Sudakov logarithms. Decomposing the dipole $S$-matrix into its real and imaginary parts, $S_x(\mathbf r) = P_x(|\mathbf r|) + i\epsilon_\perp^{\mu\nu} r_\mu S_{\perp\nu} M_p\, O^\perp_{1T,x}(|\mathbf r|)$, where $P_x$ is the pomeron and $O^\perp_{1T,x}$ the spin-dependent odderon amplitude, one finds~\cite{Bhattacharya:2025bqa}
\begin{align}
f_{q/p}\left(x,b,\mu_{i},\zeta_{i}\right) &=\frac{N_{c}B_{\perp}}{4\pi^4}\frac{1}{x}\int \dd[2]{\mathbf{r}} \int_0^\infty \dd{\epsilon_{f}^{2}}\,\mathcal{K}\left(\mathbf{b},\mathbf{r},\epsilon_f\right) \nn\\
&\quad\times \left[1 - P_x(|\mathbf r|)\right], \label{eq:unpol_x_expanded}\\
f_{1T,q/p}^{\perp}\left(x,b,\mu_{i},\zeta_{i}\right) &=\frac{N_{c}B_{\perp}}{4\pi^4}\frac{1}{x}\int \dd[2]{\mathbf{r}} \int_0^\infty \dd{\epsilon_{f}^{2}}\,\mathcal{K}\left(\mathbf{b},\mathbf{r},\epsilon_f\right) \nn\\
&\quad\times \frac{\mathbf{b} \vdot \mathbf{r} }{\mathbf{b}^2}\, O^{\perp}_{1T,x}\qty(\abs{\mathbf{r}})\,, \label{eq:Sivers_x_expanded}
\end{align}
with the common kernel
\begin{align}
\label{eq:kernel}
\mathcal{K}\left(\mathbf{b},\mathbf{r},\epsilon_f\right) = \frac{(\mathbf b + \mathbf r)\vdot \mathbf r}{|\mathbf b + \mathbf r||\mathbf r|}\, \epsilon_f^2\, K_1(\epsilon_f|\mathbf b+\mathbf r|)\, K_1(\epsilon_f |\mathbf r|)\,.
\end{align}
Here $K_1$ is the modified Bessel function of the second kind, $N_c$ is the number of colors, and $B_\perp$ the average transverse area of the proton. Equations~\eqref{eq:unpol_x_expanded} and~\eqref{eq:Sivers_x_expanded} make the C-parity assignments in Eq.~\eqref{eq:Ctmd} manifest: $P_x$ is even and the full spin-dependent term $\epsilon_\perp^{\mu\nu}r_\mu S_{\perp\nu}O^\perp_{1T,x}(|\mathbf r|)$ is odd under $\mathbf r\to -\mathbf r$, which exchanges quark and antiquark.

{\it \textbf{Numerical results.}}
We now present illustrative numerical estimates for the OPCC Sivers asymmetry. We evaluate Eq.~\eqref{eq:siversdef} using the small-$x$ expressions in Eqs.~\eqref{eq:unpol_x_expanded} and~\eqref{eq:Sivers_x_expanded}, together with the charge- and energy-weighted jet functions in Eq.~\eqref{eq:Jetfunc}. We consider the EIC kinematic setup with $Q^2=20~{\rm GeV}^2$ and $\sqrt{s}=105~{\rm GeV}$, and evaluate the asymmetry for $x\in[0.002, 0.01]$. We work at next-to-leading logarithmic (NLL) accuracy in the TMD evolution equations and take $n_f=3$ active quark flavors throughout. 

To model the dipole amplitude we specify an initial condition at $x_0 = 0.01$ and evolve it to smaller $x$ with the Balitsky--Kovchegov (BK) equation~\cite{Balitsky:1995ub,Kovchegov:1999yj}. For the pomeron we use the MV$^e$ parametrization $P_{x_0}(r) = \exp[-\frac{r^2Q_{s0}^2}{4}\ln(\frac{1}{r\Lambda_{\rm QCD}} + e_c\cdot e)]$~\cite{Lappi:2013zma}, and for the spin-dependent odderon the model of Refs.~\cite{Yao:2018vcg,Lappi:2016gqe},
\begin{equation}
\label{eq:odderonIC}
      O^{\perp}_{1T,x_0}(r) = - \kappa\, P_{x_0}\qty(r) \frac{r^2Q_{s0}^{3}}{8 M_p}\,,
\end{equation}
where we choose $\kappa=1/3$~\cite{Bhattacharya:2025bqa} as an illustrative benchmark for the odderon normalization, which is not fixed by data. The BK equation is solved with a running-coupling, using $n_f = 3$ and $\Lambda_{\rm QCD} = 0.241~{\rm GeV}$, and the initial-condition parameters are taken from the Bayesian fit to HERA data of Ref.~\cite{Casuga:2023dcf}: $Q_{s0}^2=0.061\,\mathrm{GeV}^2$, $C^2 = 4.97$, $e_c = 35.3$, and $B_\perp = 14.1~{\rm mb}$.

The non-perturbative large-$b$ region is treated with the $b^\ast$ prescription~\cite{Collins:1984kg}, $b^\ast = b/\sqrt{1+b^2/b_{\rm max}^2}$ with $b_{\mathrm{max}}=1.5\,\mathrm{GeV}^{-1}$. All TMDs are evaluated at their natural initial scales $\mu_i = \sqrt{\zeta_i} = \mu_{b^\ast} = 2e^{-\gamma_E}/b^\ast$ and evolved to the hard scales $\mu=\sqrt\zeta = Q$, which produces the Sudakov suppression at large $b$.

For the jet functions in Eq.~\eqref{eq:Jetfunc}, we write the TMD FFs as~\cite{Echevarria:2020hpy,Sun:2014dqm}
\begin{align}
    D_{h/q}\left(z,b,\mu,\zeta\right)  = D_{h/q}\left(z, b^\ast ,\mu_i,\zeta_i\right) 
e^{-S^{D}_{\text{NP}}-S_{\text{pert}}\left(\mu,\mu_i,\zeta,\zeta_i\right)}\,,\label{eq:EvolvedFF}
\end{align}
where $D_{h/q}(z,b^*,\mu_i,\zeta_i)$ is matched onto the collinear DSS charged-hadron FFs~\cite{Borsa:2023zxk} with one-loop matching coefficients, $S_{\text{pert}}$ is the NLL perturbative Sudakov factor~\cite{Echevarria:2020hpy}, and 
\begin{align}
S^{D}_{\text{NP}}\left(z,b,Q_{0},\zeta\right)=\frac{g_{2}}{2}\ln{\left(\frac{b}{b^*}\right)}\ln{\qty(\frac{\sqrt{\zeta}}{Q_{0}})}+g_{1}^{D}\frac{b^{2}}{z^{2}}\,.\label{eq:NPSudakov}
\end{align}
with $g_2 = 0.84$, $Q_0^2 = 2.4~\mathrm{GeV}^2$, and $g_1^D=0.042~{\rm GeV}^2$~\cite{Sun:2014dqm}. The same FFs and the same non-perturbative parameters enter the charge-weighted and energy-weighted jet functions; only the weight in the integrand differs. We take $z_{\min} = 0.01$ in both jet functions.

For the small-$x$ Sivers function in Eqs.~\eqref{eq:TMDpolfac} and \eqref{eq:Sivers_x_expanded}, after evolution, we have
\begin{align}
    f_{1T,q/p}^\perp\left(x,b,\mu,\zeta\right)  =& f_{1T,q/p}^\perp\left(x,b^\ast ,\mu_i,\zeta_i\right) \nn \\
&\times e^{-S^{F}_{\text{NP}}-S_{\text{pert}}\left(\mu,\mu_i,\zeta,\zeta_i\right)}\,,\label{eq:Evolved jetfunction}
\end{align}
where $S^F_{\rm NP}$ is the non-perturbative Sudakov factor. For  small-$x$ quark TMDs, we follow Ref.~\cite{Bhattacharya:2025bqa} and retain in $S_{\rm NP}^{F}$ only the contribution from the non-perturbative part of the Collins--Soper kernel, given by the term proportional to $g_2$ in Eq.~\eqref{eq:NPSudakov}. The intrinsic transverse momentum dependence of the incoming proton is assumed to be already encoded in the dipole amplitude.

\begin{figure}[t]
\begin{center}
\includegraphics[width=0.9\columnwidth]{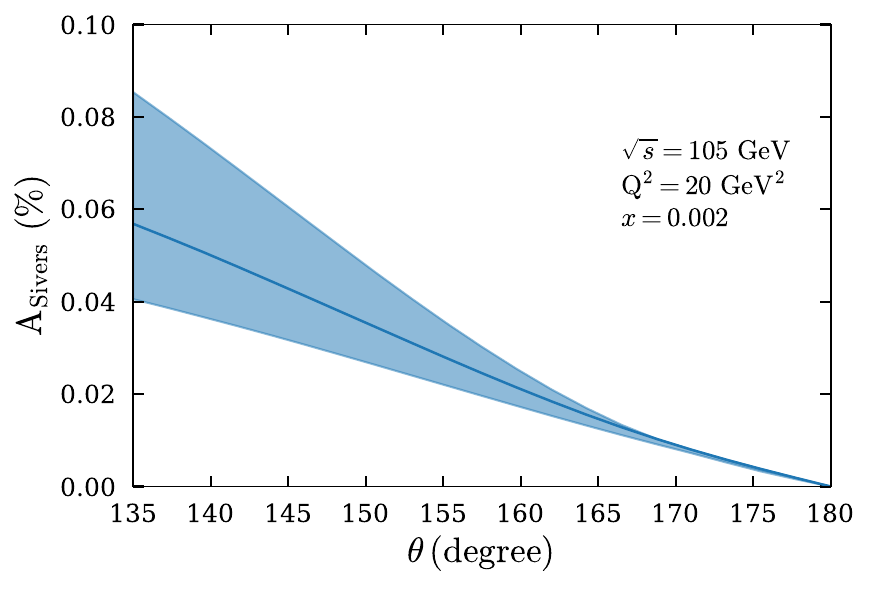}
\caption{OPCC Sivers asymmetry $A_{\rm Sivers}(\theta,x)$ of Eq.~\eqref{eq:siversdef} as a function of $\theta$ at $Q^2=20~{\rm GeV}^2$, $\sqrt{s}=105~{\rm GeV}$, and $x=0.002$, computed in the small-$x$ dipole framework. The band shows the perturbative scale variation described in the text.}
\label{fg:dFUTdtheta}
\end{center}
\end{figure}

\begin{figure}[t]
\begin{center}
\includegraphics[width=0.9\columnwidth]{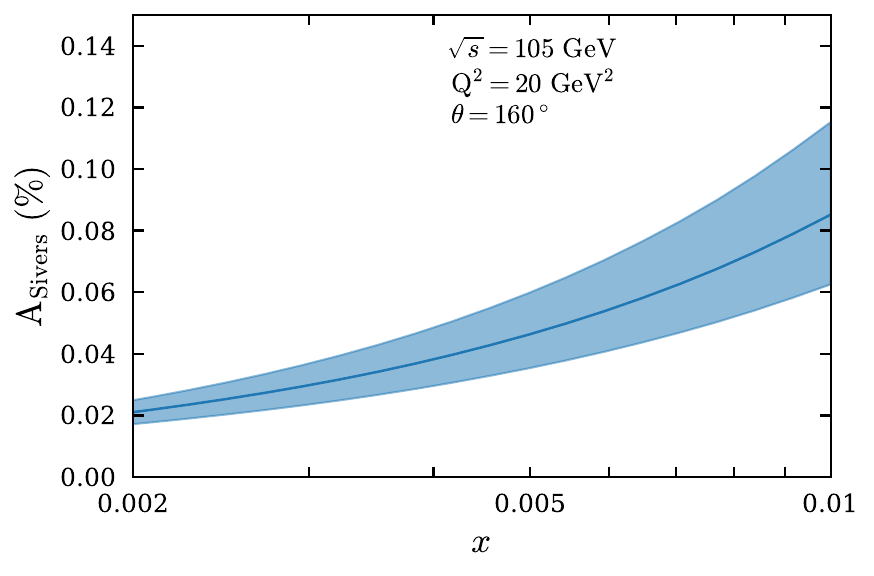}
\caption{OPCC Sivers asymmetry as a function of $x$ with
$Q^2=20~{\rm GeV}^2$, $\sqrt{s}=105~{\rm GeV}$, and $\theta=160^\circ$. The blue band shows the perturbative scale variation of the small-$x$ dipole prediction, as described in the text.}
\label{fg:dFUTdx}
\end{center}
\end{figure}

In Fig.~\ref{fg:dFUTdtheta}, we show the Sivers asymmetry as a function of $\theta$ at $x=0.002$. The band shows only the perturbative scale uncertainty, estimated by varying $\sqrt{\zeta}$ and $\mu$ by a factor of two around $Q$. The asymmetry increases as the measured direction moves away from the exact back-to-back configuration $\theta = 180^\circ$, because the $J_1$ Bessel function in the numerator vanishes at $\bar\theta = 0$ while the $J_0$ in the denominator does not, and it reaches approximately $0.05\%$ at $\theta = 140^\circ$. Since the spin-dependent odderon is not constrained, the size of the asymmetry is only for illustration based on our model. With our chosen positive normalization $\kappa=1/3$ in Eq.~\eqref{eq:odderonIC}, the model yields a positive asymmetry in the displayed kinematics. This sign reflects our model choice; a measurement of the asymmetry would constrain the sign of the spin-dependent odderon normalization.

In Fig.~\ref{fg:dFUTdx}, we show the $x$ dependence of $A_{\rm Sivers}$ at $\theta = 160^\circ$. The explicit $1/x$ prefactors in Eqs.~\eqref{eq:unpol_x_expanded} and~\eqref{eq:Sivers_x_expanded} cancel in the ratio in Eq.~\eqref{eq:siversdef}. At fixed $Q$ and $\theta$, the remaining $x$ dependence mainly arises from BK evolution of the odderon and pomeron amplitudes. As $x$ decreases, nonlinear small-$x$ evolution generally enhances the pomeron scattering amplitude while suppressing the magnitude of the spin-dependent odderon amplitude~\cite{Motyka:2005ep,Yao:2018vcg}. This behavior explains the decrease of the asymmetry toward smaller $x$. The central value decreases from approximately $0.09\%$ at $x=0.01$ to $0.02\%$ at $x=0.002$. For the chosen odderon model, the magnitude of $A_{\rm Sivers}$ over this range is $0.02\%$--$0.09\%$. Since the odderon normalization is not constrained by data, these small values should serve as an illustrative benchmark for experimental sensitivity studies; the physical asymmetry may differ substantially. Dedicated measurements across $x$ and $\theta$ are needed to determine its magnitude and kinematic dependence.

{\it \textbf{Conclusions.}}
In this Letter, we have proposed the one-point charge correlator in transversely polarized DIS as a novel probe of the spin-dependent odderon at small-$x$. We developed the theoretical formalism for the OPCC Sivers asymmetry by combining TMD factorization with the small-$x$ dipole description, expressing its charge-weighted numerator and energy-weighted denominator in terms of the spin-dependent odderon and pomeron amplitudes, respectively.

A central prediction is that, in the small-$x$ eikonal limit, the C-even pomeron contribution to the unpolarized OPCC vanishes identically, while the spin-dependent OPCC survives and is doubled by the addition of the antiquark channel. This follows from the charge weight and the odderon-generated sea-quark Sivers function both being odd under charge conjugation. The resulting asymmetry retains sensitivity to the dipole amplitudes, their evolution, and fragmentation input.

Our model estimates of the Sivers asymmetry at Electron-Ion Collider kinematics provide illustrative guidance for experimental studies. Since the spin-dependent odderon remains unconstrained by data, we encourage dedicated measurements of the OPCC asymmetry over a range of $x$ and $\theta$ to constrain the nonperturbative initial condition of the spin-dependent odderon amplitude, including its sign and normalization. Experimentally, the hadronic information entering both the numerator and the denominator is obtained entirely from charged-particle tracks. The OPCC thus provides a direct and complementary avenue for probing the spin-dependent odderon and, more broadly, the quark Sivers function in the small-$x$ regime.

{\it \textbf{Acknowledgment.}} We thank Frank Petriello and Zhen Xu for useful discussions. Z.K., D.P., and J.P. are supported by the National Science Foundation under grant No.~PHY-2515057. This work is also supported by the U.S. Department of Energy, Office of Science, Office of Nuclear Physics, within the framework of the Saturated Glue (SURGE) Topical Theory Collaboration. H.C. is supported by the U.S. Department of Energy, Office of High Energy Physics, under contract No.~DE-SC0010143, and is partially supported by a CFNS Joint Postdoctoral Fellowship. This research was supported in part through the computational resources and staff contributions provided for the Quest high performance computing facility at Northwestern University which is jointly supported by the Office of the Provost, the Office for Research, and Northwestern University Information Technology. Z.K. would like to thank the Erwin-Schr\"odinger International Institute for Mathematics and Physics at the University of Vienna for partial support during the Programme `New Paradigms for Harnessing Quantum Field Theory at Colliders', July 27 – August 28, 2026.

\FloatBarrier 

\bibliographystyle{JHEP-2modlong.bst}
\bibliography{references}

\end{document}